\documentstyle[twocolumn,seceq]{jpsj}
\renewcommand{\vec}[1]{\mbox{\boldmath $#1$}}

\def\runtitle{Competition between Hund-Rule Coupling and Kondo Effect}
\def\runauthor{Hiroaki {\sc Kusunose} and Kazumasa {\sc Miyake}}

\title
{
Competition between Hund-Rule Coupling and Kondo Effect
}

\author
{
Hiroaki {\sc Kusunose} and Kazumasa {\sc Miyake}
}

\inst
{
Department of Material Physics, Faculty of Engineering Science,\\
Osaka University, Toyonaka 560
}

\recdate
{
\today
}

\abst
{
We investigate a problem about the competition between the hybridization and
the Hund-rule coupling by applying the Wilson numerical renormalization-group
method to the extended Kondo model where the impurity spin interacts
via the Hund-rule coupling, with an extra spin which is isolated from the
conduction electrons.
It is shown that the Hund-rule coupling is an irrelevant perturbation against the strong coupling fixed point.
However, the Hund-rule coupling decreases the characteristic energy $T_{\rm K}$
drastically to the lower side and the irrelevant operator, which describes the low energy
physics, takes a form of {\it ferromagnetic} exchange interaction between the
extra spin and the Kondo resonance states because of the existence of
the Hund-rule coupling.
}

\kword
{
Hund-rule coupling, Kondo effect, numerical renormalization-group,
anisotropic hybridization, plural electrons at localized orbitals,
transition between high-spin and low-spin state,
Uranium based heavy fermion, Ni-doped High-$T_{\rm c}$ cuprate
}

\begin{document}
\sloppy
\maketitle

\section{Introduction}
One of the most important question concerning heavy fermion systems\cite{grewe}, which
exhibit the exotic phenomena such as anisotropic superconductivity and
extremely weak anitiferromagnetism is to understand how the
low energy quasipartile states are formed, in other words, how the high energy
incoherent states are reflected in the low energy physics.
Many theoretical attempts have been made to include higher Crystalline Electric
Field (CEF) effects both in single
impurity\cite{cox,kuramoto,koga,tskim} and lattice
case\cite{evans,trees,anders}. When the degenerate orbitals are
deformed by CEF, the hybridizations between the deformed orbitals and conduction
electrons become anisotropic in general.
The anisotropic hybridization produces different characteristic energies for
each CEF orbitals. As a result, the orbital which concerns the Kondo effect changes
depending on the temperature range.

For Ce based compounds this anisotropy is reflected in the formation of highly renormalized quasiparticle band only through the one-body effect:
its typical example has been put forward in Ref. 9 to explain anomalous properties observed in CeNiSn.
On the other hand, for U based compounds, where $(5f)^2$ or $(5f)^3$
configuration is realized, the anisotropic hybridizations may be reflected on
the quasiparticles through a many-body effect, because at least two
characteristic energy scales and the Hund-rule coupling are
involved in the problem.
The ``spin of localized electron" tends to be quenched by the Kondo effect,
while the Hund-rule coupling stabilizes the high-spin state.
Then a competition between the two effects plays a crucial role in determining the low energy physics.

Such a competition is likely to be realized in a variety of physical situations:
For example, (i) the magnetic susceptibility in UPd$_2$Al$_3$ can be fitted by
a certain CEF scheme under the tetravalent U state
($(5f)^2$ configuration)\cite{grauel} while the Fermi surface
measured by dHvA effect is in good agreement with the band structure
calculation with the trivalent state ($(5f)^3$ configuration)\cite{band},
and the photoemission data are reported to be explained by $(5f)^3$ configuration\cite{takahashi}.
(ii) in Ni-doped High-$T_{\rm c}$ cuprates, the spin moment of Ni$^{2+}$
changes from the high-spin to low-spin state with increasing the doping rate
of holes\cite{mendels,kodama,zheng,sera,kuroda}, and so on.

In this paper, we discuss how the competition between the effect of anisotropic hybridization and the Hund-rule coupling affects the Kondo screening on the
basis of the minimal model including these features.
In \S 2, starting from an extended Anderson model, we derive the usual Kondo
exchange model with the extra spin interacting via the Hund-rule coupling.
In \S 3, we discuss the effect of the Hund-rule coupling on the Kondo effect of
the model derived in \S 2 by means of the numerical renormalization-group
method\cite{wilson,wilkins}.
In the final section, we summarize the results and discuss their implications
to the realistic problem.

\section{Model}
When we discuss the behavior of magnetic impurity in real metals,
we have to deal with a degenerate Anderson model including the
Hund-rule coupling and CEF as well as the direct Coulomb interaction.
In addition to this, we also have to take into account the anisotropy of
hybridization specific to each orbitals split by CEF.

In this paper, we make the following simplification to a generalized
Anderson model.
\begin{itemize}
\item The effect of CEF is implicitly included in such a way that the
hybridizations for each orbitals are different according to the guide of the
point-group theory.
\item The energy levels of those orbitals are assumed to be the same for
simplicity.
\item Of these orbitals only the two orbitals are retained.
\item The hybridization for one orbital is taken finite, while that for the
other one is neglected as the limiting case for different hybridizations.
\item Both the Hund-rule coupling $J_{\rm H}$ and the direct Coulomb
interactions $U$, which are independent of orbitals, are retained.
\end{itemize}
Thus, we are left with an extended Anderson model as,
\begin{eqnarray}
\label{eq:001}
&& H = \sum_{k\sigma} \epsilon_{k}a^\dagger_{k\sigma}a_{k\sigma}
+ H_{\rm mix} + H_f, \\
\label{eq:002}
&& H_{\rm mix} = \sum_{k\sigma}(V_{k}a^\dagger_{k\sigma}f_{1\sigma}+
{\rm h.c.}), \\
\label{eq:003}
&& H_f = E_f\sum_{m\sigma}f^\dagger_{m\sigma}f_{m\sigma}
+ \frac{U}{2}\sum_{mm'}\sum_{\sigma\sigma'}
f^\dagger_{m\sigma}f^\dagger_{m'\sigma'}f_{m'\sigma'}f_{m\sigma}\nonumber\\
&&\mbox{\hspace{1cm}}+ \frac{J_{\rm H}}{2}\sum_{mm'}\sum_{\sigma\sigma'}
f^\dagger_{m\sigma}f^\dagger_{m'\sigma'}f_{m\sigma'}f_{m'\sigma},
\end{eqnarray}
where $f^\dagger_{m\sigma}$ denotes the creation operator for the localized
electron with spin $\sigma$ ($=\uparrow,\downarrow$) on the orbital $m$ ($=1,2$)
deformed by the CEF, and $a^\dagger_{k\sigma}$ for
conduction electron with the wave number $k$ and the spin $\sigma$ on the band
hybridizing only with the localized orbital ($m = 1$) via the hybridization
$V_{k}$.

We can rewrite the local part $H_f$, (\ref{eq:003}), of the Hamiltonian as
\begin{equation}
\label{eq:004}
H_f = E_fn_f+\frac{U}{2}(n_f^2-n_f)-J_H(\vec{S}_f^2+\frac{1}{4}n_f^2-n_f)
\end{equation}
in terms of the number and the total spin of localized electrons defined by
\begin{eqnarray}
\label{eq:005}
&&n_f = \sum_m n_{mf}=\sum_{m}\sum_{\sigma}f^\dagger_{m\sigma}f_{m\sigma},\\
\label{eq:006}
&&\vec{S}_f = \sum_{m}\vec{S}_{mf}=\sum_m
\frac{1}{2}\sum_{\sigma\sigma'}f^\dagger_{m\sigma}\vec{\sigma}_{\sigma\sigma'}
f_{m\sigma'},
\end{eqnarray}
where $\vec{\sigma}$ is the vector of the Pauli matrices.

In order to simplify this model furthermore, we restrict our discussions within
the case of strong Coulomb interaction. Then we can treat the hybridization
term $H_{\rm mix}$ in eq. (\ref{eq:001}) within the second order pertubation
by restricting the Hilbert space in such a way that the number of localized
electrons on ground state is given by $\langle n_f\rangle=2$.
Such a restriction is valid when $(E_f+U)(E_f+2U)<0$ and $E_f,U\gg J_{\rm H}$.
Thus we can get the effective Hamiltonian (as shown in Appendix):
\begin{eqnarray}
\label{eq:007}
&&H = \sum_{k\sigma}\epsilon_{k}a^\dagger_{k\sigma}a_{k\sigma}
-J_{\rm H}(\vec{S}_f^2+\frac{1}{4}n_f^2-n_f)\nonumber\\
&&\mbox{\hspace{0.2cm}}+ J\sum_{\rm f,i}|{\rm f}\rangle\langle{\rm i}|\langle {\rm f}|\vec{S}_{1f}|{\rm i}\rangle\cdot
\sum_{k\sigma}\sum_{k'\sigma'}
a^\dagger_{k\sigma}\vec{\sigma}_{\sigma\sigma'}a_{k'\sigma'},
\end{eqnarray}
where $|{\rm i}\rangle$ and $|{\rm f}\rangle$ are initial and final states of the localized
electrons and the exchange coupling $J$ is given by
\begin{equation}
\label{eq:008}
J = \left[\frac{1}{E_f+2U}-\frac{1}{E_f+U}\right]|V_{k_{\rm F}}|^2
> 0.
\end{equation}

Now we consider the two limiting cases, (i) $J_{\rm H}\ll J$ and (ii)
$J_{\rm H}\gg J$, in which the second term in (\ref{eq:007}), the Hund-rule coupling, takes a simple form.

\begin{flushleft}(i) $J_{\rm H}\ll J$:\end{flushleft}

We can treat the Hund-rule coupling as a perturbation on the Hamiltonian
eq. (\ref{eq:007}) with $J_{\rm H}=0$. We can show the following identities
valid in the restricted Hilbert space such that $\langle n_{1f}\rangle=\langle n_{2f}\rangle=1$,
\begin{eqnarray}
\label{eq:009}
&&\sum_{\rm i,f}|{\rm f}\rangle\langle {\rm i}|\langle {\rm f}|\vec{S}_{1f}|{\rm i}\rangle = \vec{S} \\
\label{eq:010}
&&-J_{\rm H}\left(\vec{S}_f^2+\frac{1}{4}n_f^2-n_f\right)
=-2J_{\rm H}\vec{S}\cdot\vec{S}'-\frac{J_{\rm H}}{2},
\end{eqnarray}
where $\vec{S}$ and $\vec{S}'$ are the spin matrices of $S=S'=1/2$,
respectively.
Thus, we get the $S=1/2$ $s$-$d$ exchange Hamiltonian interacting with a spin
of $S'=1/2$ via the Hund-rule coupling:
\begin{equation}
\label{eq:011}
H = \sum_{k\sigma}\epsilon_{k}a^\dagger_{k\sigma}a_{k\sigma}
+J\vec{S}\cdot\sum_{k\sigma}\sum_{k'\sigma'}
a^\dagger_{k\sigma}\vec{\sigma}_{\sigma\sigma'}a_{k'\sigma'}
-2J_{\rm H}\vec{S}\cdot\vec{S}'.
\end{equation}

\begin{flushleft}(ii) $J_{\rm H}\gg J$:\end{flushleft}

In this case, the spin triplet state is realized as a local ground state,
and then the matrix element for the localized electron, interacting with
conduction electrons, can be related to the spin of $S=1$ as
\begin{equation}
\label{eq:012}
\sum_{\rm i,f}|{\rm f}\rangle\langle {\rm i}|\langle {\rm f}|\vec{S}_{1f}|{\rm i}\rangle = \frac{1}{2}\vec{S}.
\end{equation}
Thus we can get the $S=1$ $s$-$d$ exchange Hamiltonian as a model in this limit:
\begin{equation}
\label{eq:013}
H = \sum_{k\sigma}\epsilon_{k}a^\dagger_{k\sigma}a_{k\sigma}
+\frac{J}{2}\vec{S}\cdot\sum_{k\sigma}\sum_{k'\sigma'}
a^\dagger_{k\sigma}\vec{\sigma}_{\sigma\sigma'}a_{k'\sigma'}.
\end{equation}
It is noted that the exchange coupling constant is devided by a factor
$2$ because only the half of spin $\vec{S}$, i.e., $\vec{S}_{1f}$,
interacts with the conduction electrons.
Therefore the characteristic energy $T_{\rm K}/D\sim e^{-2/\rho_{\rm F}J}$
for this model is far less than $T_{\rm K}^0/D\sim e^{-1/\rho_{\rm F}J}$
for a usual $S=1$ exchange model
unless $T_{\rm K}^0$ is comparable to $D$. Here, $D$ is half of the bandwidth
and $\rho_{\rm F}$ the density of states of conduction electrons at the
Fermi level.

\section{Effect of Hund-rule coupling}
We investigate the truncated model (\ref{eq:011}) in order to study the effect
of the Hund-rule coupling on the Kondo effect.
We put no restriction on the couplings $J$ and $J_{\rm H}$, despite the model is
derived under the condition $J_{\rm H}\ll J$.
Practically, we calculate the RG flow of excitation energies and the
temperature dependence of the magnetic susceptibility for the impurity spin by
using the Wilson numerical renormalization-group (NRG) method\cite{wilson,wilkins}.

In order to use NRG method, we translate the kinetic energy part in the
Hamiltonian (\ref{eq:011}) into a tridiagonal form as usual:
\begin{eqnarray}
\label{eq:014}
&&H_N=\Lambda^{(N-1)/2}\left[
\sum_{n=0}^{N-1}\sum_{\sigma}\Lambda^{-n/2}(f^\dagger_{n\sigma}f_{n+1\sigma}
+{\rm h.c.})\right.\nonumber\\
&&\mbox{\hspace{1cm}}\left.+J\vec{S}\cdot\sum_{\sigma\sigma'} f^\dagger_{0\sigma}
\vec{\sigma}_{\sigma\sigma'}f_{0\sigma'}-J_{\rm H}\vec{S}\cdot\vec{S}'
\right],
\end{eqnarray}
where $f^\dagger_{n\sigma}$ is the creation operator for the conduction
electron of the Wannier representation with radial extent
$k_{\rm F}^{-1}\Lambda^{n/2}$, and the energy scales are expanded by a factor
$\Lambda^{(N-1)/2}$ and remeasuared in a unit of $(1+\Lambda^{-1})D/2$.
We use $\Lambda=2$ throughout this paper.

The flow of excitation energies is shown in Fig. 1 for $J=0.4$ and
$J_{\rm H}=0.04$.
The labels attached at the right side of each lines denote the states with the
number of particle $Q$ and the total spin $S$ as shown in the figure.
The excitation energies are measured from that of the state, $e_0$ for
even iterations and $o_1$ for odd iterations, as it will turn out
to be convenient later.

It is remarked that the flow lines gradually converge to those at the fixed
point. We can reproduce the level structure of the excitation energies at
the fixed point by combining with one-particle excitations which are
determined by the exchange Hamiltonian (\ref{eq:014}) with $J = \infty$ and
$J_{\rm H}=0$; in other words, the couplings $J$ and $J_{\rm H}$ approach
$J^*=\infty$ and $J_{\rm H}^*=0$, respectively, as the renormalization step
proceeds.
In order to see this, we plot the flow lines denoted by dots for $J=20.0$ and
$J_{\rm H}=0$.

The tendency that the exchange coupling is renormalized to the strong coupling
one and the Hund-rule coupling is weaken can be seen also by the perturbational
renormalization-group argument. Indeed, the lowest non-vanishing
renormalization of $J$ and $J_{\rm H}$ are given by the scaling equations
\begin{eqnarray}
&&\frac{{\rm d}J}{{\rm d}\ln(D/D_0)}=-J^2,\\
&&\frac{{\rm d}J_{\rm H}}{{\rm d}\ln(D/D_0)}=J^2J_{\rm H},
\end{eqnarray}
which arises from the processes shown in Fig. 2.
Solutions of eqs. (3.2) and (3.3) are given by
\begin{eqnarray}
&&J=\frac{J_0}{1+J_0\ln(D/D_0)},\\
&&J_{\rm H}=J_{\rm H}^0e^{-(J-J_0)}.
\end{eqnarray}
As a result, the exchange coupling $J$ approaches the strong coupling and
the Hund-rule coupling $J_{\rm H}$ is renormalized rapidly downward
simultaneously.

The way to approach the fixed point is quite gradual as compared with
the way in the case of $S=1/2$, because the irrelevant operator consists
dominantly of the ferromagnetic exchange interaction due to the presence of
the Hund-rule coupling.
Actually, we can describe the excitations near the fixed point by the
effective Hamiltonian with the ferromagnetic exchange interaction,
\begin{eqnarray}
\label{eq:015}
&&H_N=\Lambda^{(N-1)/2}\left[
\sum_{n=1}^{N-1}\sum_{\sigma}\Lambda^{-n/2}(f^{*\dagger}_{n\sigma}f^*_{n+1\sigma}+
{\rm h.c.})\right.\nonumber\\
&&\mbox{\hspace{1cm}}\left.-J_{\rm eff}(N)\vec{S}'\cdot\sum_{\sigma\sigma'} f^{*\dagger}_{1\sigma}
\vec{\sigma}_{\sigma\sigma'}f^*_{1\sigma'}\right],
\end{eqnarray}
where $f^*_{n\sigma}$ represents the quasiparticle associated with Kondo
resonance state, reflecting the singlet formation between the spin $S$ and the conduction electrons.
It is noted that the spin $\vec{S}'$, which is left after the spin $\vec{S}$
is compensated by the conduction electrons, couples with the conduction
electrons at the ``site" $n=1$ because the couduction electrons at $n=0$
are wiped out due to the singlet formation with $\vec{S}$.
We can estimate the $N$-dependence of the ferromagnetic exchange coupling by
using the first order perturbation with respect to $J_{\rm eff}(N)$.
$f^*_{1\sigma}$ is expressed in terms of the eigenstates $g_l$ and $h_l^\dagger$ of
the $N$-site free chain Hamiltonian, the first term in eq. (\ref{eq:015}), as
\begin{equation}
f^*_{1\sigma}=\left\{
\begin{array}{l}
\Lambda^{-(N-1)/4}\sum_{l=1}^{N/2}\alpha_l(g_l+h^\dagger_l)\\
\Lambda^{-(N-1)/4}\left[\sum_{l=1}^{(N-1)/2}\alpha_l'(g_l+h_l^\dagger)+\alpha_0'g_0\right],
\end{array}\right.
\end{equation}
for even and odd $N$ respectively, where $\alpha_l$ and $\alpha'_l$ for $\Lambda=2$ are obtained by the numerical
diagonalization as
\begin{eqnarray}
\alpha_1=0.6998,\;\;\;\alpha_2=0.7480,\;\;\;\ldots,\\
\alpha'_0=0.7071,\;\;\;\alpha'_1=0.6980,\;\;\;\ldots.
\end{eqnarray}
Since the exchage interaction has an eigenvalue $\Lambda^0$ with respect to
the linearized renormalization-group transformation, the operator itself is
marginal.
As we can see below, $J_{\rm eff}$ is renormalized to zero as similar to
the case of the weak coupling fixed point in the usual ferromagnetic Kondo
problem.
The first excitation energies are estimated within the first order perturbation
as $2J_{\rm eff}(N)\alpha_1^2$ for even iterations and
$2J_{\rm eff}(N)\alpha'^2_0$ for odd iterations.
Thus, by comparing these results with those shown in Fig. 1, we can estimate
the $N$-dependence of $J_{\rm eff}$ as shown in Fig. 3. This $N$-dependence
can be fitted by the well-known form in the ferromagnetic exchange Kondo
problem given by
\begin{equation}
\label{eq:016}
J_{\rm eff}(N) = \frac{\alpha J_0}{1+J_0\ln\Lambda(N-N_0)/2},
\end{equation}
where we put $T/D=\Lambda^{-(N-1)/2}$ and fitting parameters are put
$\alpha=1.306$, $J_0=0.3027$ and $N_0=20$, respectively. We also confirm
that the higher excitations can be fitted by means of the effective exchange
$J_{\rm eff}(N)$.

This result implies that $J$ and $J_{\rm H}$ are renormalized as
$J\rightarrow\infty$ and $J_{\rm H}\rightarrow 0$ even before $N\sim 25$
iteration and the effective ferromagnetic exchange intraction $J_{\rm eff}(N)$
between the conduction electrons, and after $N\sim 25$ iteration the spin $\vec{S}'$ rules the excitations as the irrelevant operator around the fixed point $J=\infty$, $J_{\rm H}=0$, and $J_{\rm eff}(N)=0$.
This is understood as follows. When $J(T)\gg J_{\rm H}(T)$, ``local" spin
$\vec{S}$ forms singlet with ``0th" conduction electron at ``0th" site and
there are two degeneracies with respect to the degrees of freedom of the
extra spin $\vec{S}'$. If the virtual hopping process between the ``1st" and
``0th" site is taken into account, the renormalized Hund-rule coupling
$J_{\rm H}(T)$ lifts the degeneracy leaving the state, where
the spin $\vec{S}'$ and the spin of conduction electron at the ``1st" site
are parallel, be lower energy than that of anti-parallel.
This mechanism is similar to that discussed in Ref. 20 for the origin of
anomaly of multichannel Kondo effect.

Next, we discuss the temperature dependence of the susceptibility for the
impurity spin which is defined by
\begin{equation}
\label{eq:017}
T\chi_{\rm imp}=\lim_{N\rightarrow\infty}
\left[
\frac{{\rm Tr}S_{z,N}^2{\rm e}^{-\beta_NH_N}}{{\rm Tr e}^{-\beta_NH_N}}
-\frac{{\rm Tr}S_{z,N}^2{\rm e}^{-\beta_NH_N^0}}{{\rm Tr e}^{-\beta_NH_N^0}}
\right],
\end{equation}
where $S_{z,N}$ is the $z$-component of the total spin
\begin{equation}
S_{z,N}=S_z+S'_z+\sum_{n=0}^{N}\sum_{\sigma}\frac{1}{2}f^\dagger_{n\sigma}
\sigma_{\sigma\sigma}^zf_{n\sigma},
\end{equation}
and the second term of (\ref{eq:017}) represents the contribution of non-interacting system, and $\beta_N$ is defined as
\begin{equation}
\label{eq:018}
\beta_N=\Lambda^{-(N-1)/2}/T.
\end{equation}
By setting $T=T_N\equiv\Lambda^{-(N-1)/2}$, we can determine the susceptibility
with a good accuracy because the excited states with the energy
$\beta_N^{-1}\sim 1$, which contribute dominantly to the thermodynamic
quantity, are obtained with a good accuracy in NRG calculation\cite{wilkins}.

The temperature dependence of the susceptibility $\chi_{\rm imp}$ is shown
in Fig. 4. One can see that the Hund-rule coupling makes the characteristic
energy, $T_{\rm K}$, be lower and the $T$-dependence of $T\chi_{\rm imp}$
in the limit of $J_{\rm H}\rightarrow\infty$ is equivalent to that of
$S=1$ exchange model,
eq. (\ref{eq:013}), with $J/2=0.2$, half of $J=0.4$ in the model (\ref{eq:014}).
It is important to note that the characteristic energy $T_{\rm K}$ decreases
remarkably by three orders of magnitude even if $J_{\rm H}\sim J$.
$T\chi_{\rm imp}$ approaches toward $1/4$, the value for the case of
free spin $S=1/2$, as temperature decreases well below $T_{\rm K}$. This is
consistent with the fact that the fixed point is given by $J^*=\infty$ and
$J_{\rm H}^*=0$.
It is remarked that the fixed point remains the same as the case in the
absence of the Hund-rule coupling while the Hund-rule coupling changes
$T_{\rm K}$ drastically to the lower value.
The way to approach the fixed point becomes gradual with the
increase of $J_{\rm H}$ as shown in Fig. 4. In the case $J_{\rm H}=0$,
$T\chi_{\rm imp}$ obviously approaches the fixed point in the same way
as the conventional antiferromagnetic Kondo effect, apart from the contribution
of extra spin $\vec{S}'$, i.e., $1/4$.
In the case $J_{\rm H}\gg J$, on the other hand, the temperature
dependence of $T\chi_{\rm imp}$ becomes coincident with that of $S=1$ Kondo
effect.

We can show that the way to approach the fixed point of $S=1$ Kondo effect
is equivalent to that of $S=1/2$ {\it ferromagnetic} Kondo effect as shown
in Fig. 5, in which the temperature dependence of $T\chi_{\rm imp}$ for
$J=0.4$ and $J_{\rm H}=0.04$ in eq. (\ref{eq:014}) is compared with the
ferromagnetic $S=1/2$ Kondo effect with $J=-7.4$.
This result shows that the ferromagnetic exchange interaction behaves as
an irrelevant operator in consistent with the previous discussions about the
excitation level scheme in the presence of the Hund-rule coupling $J_{\rm H}$.

\section{Conclusions and Discussions}
We have investigated the effect of the Hund-rule coupling by using the
extended Kondo model in which the impurity spin interacts with the conduction
electrons and the other spin, which itself is decoupled from conduction
electrons, via the Hund-rule coupling as well. We have derived this model from
the generalized Anderson model which has two orbitals, where
the one orbital has more localized character than the other and its
hybridization with the conduction electrons is neglected. It means that we
restrict our concern within the temperature range above the lower
characteristic energy corresponding to the smaller hybridization.
We have concluded that Kondo effect is always dominant even if the exchange
coupling is much smaller than the Hund-rule coupling.
However, we have shown two prominent effects of the Hund-rule coupling beyond
the conventional Kondo model.

First, the characteristic energy is drastically decreased
from the value of the usual $S=1/2$ Kondo model to that of the $S=1$ Kondo
model with half of the exchange coupling, as the Hund-rule coupling increases.
This gives us a hint on general question how
the lower characteristic energy scales are realized in uranium based heavy
fermions with the hybridization larger than that of the Ce based compounds.
Namely, if there exists a localized orbital, which less hybridizes with ligand
conduction electrons due to the symmetry reason, it can work to reduce the
characteristic energy scale of the localized orbital, which well hybridizes
with conduction electrons, through the Hund-rule coupling.

Second, the Hund-rule coupling turns into the ferromagnetic exchange
interaction between the extra spin $\vec{S}'$ and conduction electrons at
``1st" site. In other words, the quasiparticle associated with the Kondo
resonance state, consisting of the spin $S$ and the conduction electrons,
interact with $S'$ in the ferromagnetic exchange form.
This irrelevant operator does not change the fixed point but affects the way to
approach the fixed point. Namely, it takes the form of the ferromagnetic Kondo
effect as the Hund-rule coupling increases. However, such ferromagnetic Kondo
effect may not occur in real systems because of at least the following two
reasons: (i) if we take into accout the direct interaction between the
conduction electrons and the localized spin $S'$, no matter how small it is,
the antiferromagnetic Kondo effect leads to form the another Kondo resonance
state against the Hund-rule coupling.
(ii) if we consider more complicated CEF model, the internal degrees of freedom
at the impurity site cannot be described as the simple spin, e.g. in the case
of CEF singlet for $f^2$ configuration, the irrelevant operator may have the
tensor form in general\cite{koga}.

The results suggest that the Kondo resonance states are individually formed
in each channels and the interactions between them are irrelevant. This is
contrary to the picture that the high-spin state due to the Hund-rule coupling
is compensated by the conduction electrons in multi channels.
Thus, the consideration of a more realistic picture for the CEF structure and
the anisotropic hybridization will give us a more solid picture for the
quasiparticles.

\section*{Acknowledgements}
The authors would like to thank Dr. O. Narikiyo for his critical reading of
this manuscript and useful discussion.
This work is supported by the Grant-in-Aid for Scientific Research (07640477),
and Monbusho International Scientific Programs (07044078) and (06044135), and
the Grant-in-Aid for Scientific Research on Priority Areas ``Physics of
Strongly Correlated Conductors" (06244104) and ``Anomalous Metallic State
near the Mott Transition" (07237102) of Ministry of Education, Science
and Culture. One of the authors (H. K.) acknowledges the support from the
Research Fellowships of the Japan Society for the Promotion of Science for
Young Scientists.

\appendix
\section{}
Here we derive the exchange part of the effective Hamiltonian, (\ref{eq:007}),
from the extended Anderson model, (\ref{eq:001}), by the second order
perturbation with respect to $H_{\rm mix}$ in eq. (\ref{eq:001}).

The eigenstates of the local Hamiltonian $H_f$, (\ref{eq:004}), with $f^n$
configuration and total $f$ spin, $S_f$, for  have the eigen energy,
\begin{equation}
\label{eq:a01}
I(n,S_f) = E_fn+\frac{U}{2}(n^2-n)-J_{\rm H}\left[S_f(S_f+1)+\frac{1}{4}n^2-n\right].
\end{equation}
When we consider $n=2$ ground state, which is valid when $(E_f+U)(E_f+2U)<0$
and $E_f,U\gg J_{\rm H}$, the Schrieffer-Wolff transformation
gives us the following effective exchange interaction between the conduction
electrons and the ground states $|{\rm i}\rangle$, $|{\rm f}\rangle$ of localized electrons,
\begin{eqnarray}
\label{eq:a02}
&&H_{\rm ex} = \sum_{{\rm i,f}kk'\sigma\sigma'}
\sum_{r}\left[
\frac{\langle {\rm f}| f_{1\sigma} |{\rm r}\rangle\langle {\rm r}| f^\dagger_{1\sigma'} |{\rm i}\rangle}
{\epsilon-\{I(3,1/2)-I(2,S_f)-\epsilon_{k'}\}}
a^\dagger_{k\sigma}a_{k'\sigma'} \right.\nonumber \\
&&\left.
+\frac{\langle {\rm f}| f^\dagger_{1\sigma'} |{\rm r}\rangle\langle {\rm r}| f_{1\sigma} |{\rm i}\rangle}
{\epsilon-\{I(1,1/2)-I(2,S_f)+\epsilon_{k}\}}
a_{k'\sigma'}a^\dagger_{k\sigma}
\right]
|{\rm f}\rangle\langle {\rm i}|V_{k}V_{k'}^*.
\end{eqnarray}
We can take the sum over the intermediate states $|r\rangle$ as forming a
complete set. Then, with the use of the anticommutation relations for
$f_{m\sigma}$ and $a_{k\sigma}$, and the identity,
$\delta_{\sigma_1\sigma_4}\delta_{\sigma_2\sigma_3}=
(\vec{\sigma}_{\sigma_1\sigma_2}\cdot\vec{\sigma}_{\sigma_3\sigma_4}+
\delta_{\sigma_1\sigma_2}\delta_{\sigma_3\sigma_4})/2$,
(\ref{eq:a02}) can be reduced to the following form:
\begin{eqnarray}
\label{eq:a03}
&&H_{\rm ex} = J\sum_{\rm i,f}|{\rm f}\rangle\langle{\rm i}|\langle{\rm f}|\vec{S}_{1f}|{\rm i}\rangle\cdot
\sum_{k\sigma}\sum_{k'\sigma'}a^\dagger_{k\sigma}\vec{\sigma}_{\sigma\sigma'}
a_{k'\sigma'}\nonumber\\
&&\mbox{\hspace{1cm}} + V\sum_{\rm i}|{\rm i}\rangle\langle{\rm i}|\sum_{kk'\sigma}a^\dagger_{k\sigma}a_{k'\sigma}
+\Sigma_f,
\end{eqnarray}
where we have taken $\epsilon=\epsilon_k=-\epsilon_{k'}=\epsilon_{k_{\rm F}}$,
and the exchange interaction $J$, the strength of potential scattering $V$
and the self-energy $\Sigma_f$ of localized electrons are given by
\begin{eqnarray}
\label{eq:a04}
&&J = \left[\frac{1}{E_f+2U}-\frac{1}{E_f+U}\right]
|V_{k_{\rm F}}|^2, \\
\label{eq:a05}
&& V = -\frac{1}{2}\left[\frac{1}{E_f+2U}+\frac{1}{E_f+U}
\right]|V_{k_{\rm F}}|^2, \\
\label{eq:a06}
&& \Sigma_f = \sum_{k}\frac{|V_{k}|^2}{E_f+U}
\sum_{\rm i}\langle {\rm i}| n_{1f} |{\rm i}\rangle,
\end{eqnarray}
respectively. Here we have neglected $J_{\rm H}$ compared to $U$.
Ignoring the potential scattering and the self-energy,
we obtain the exchange Hamiltonian eq. (\ref{eq:007}).

\vspace{3mm}
\begin{flushleft}
\bf Figure Captions
\end{flushleft}
\vspace{1.5mm}
\begin{description}
\item[]Fig. 1 The flow of excitation energies for $J=0.4$ and $J_{\rm H}=0.04$. The
labels at the right side of each lines denote the states with the number of
particle $Q$ and the total spin $S$.
\item[]Fig. 2 The Feynman diagrams giving the lowest non-vanishing
renormalization of the Hund-rule coupling $J_{\rm H}$ and the exchange
coupling $J$. The solid line denotes the conduction electron, the broken
line the pseudo fermion representing the spin $\vec{S}$, and the dotted line
the pseudo fermion of $\vec{S'}$.
\item[]Fig. 3 $N$-dependence of effective ferromagnetic interaction between
Fermi liquid at strong fixed point and the extra spin $\vec{S}'$.
The dots are determined from the first excitation energies in Fig. 1. The solid
line represents the scaling form (3.10).
\item[]Fig. 4 Temperature dependence of the susceptibility for the impurity
spin described by (3.1) for various amount of $J_{\rm H}$ with $J=0.4$.
The solid line represents $T\chi_{\rm imp}$ for $S=1$ Kondo model, (2.13), with
$J/2=0.2$.
\item[]Fig. 5 Comparison between $T\chi_{\rm imp}$ of $S=1$
antiferromagnetic Kondo model and those of $S=1/2$ ferromagnetic Kondo model.
\end{description}

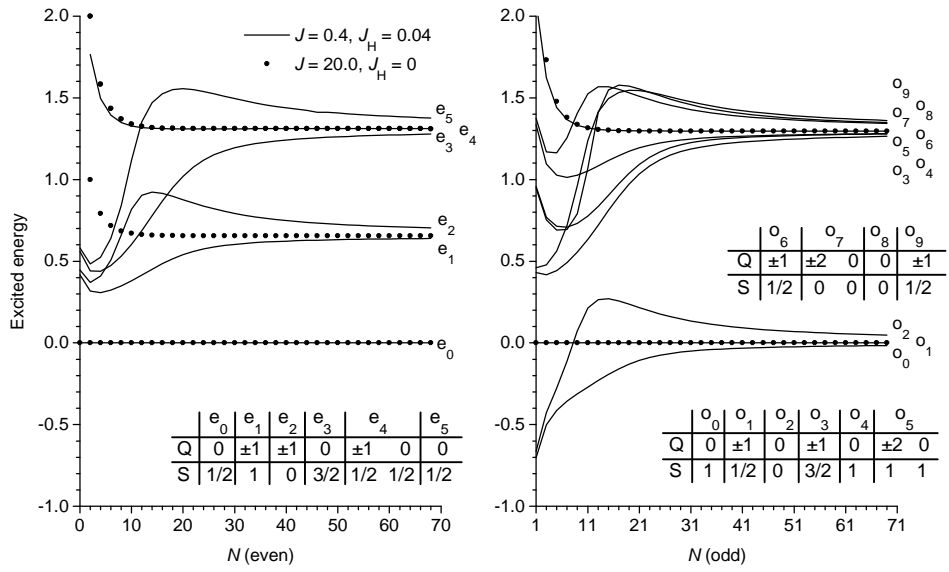
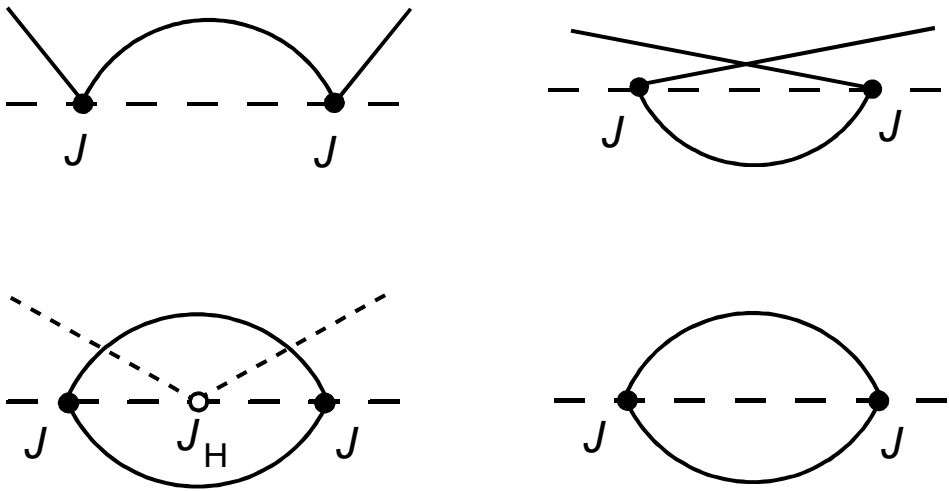
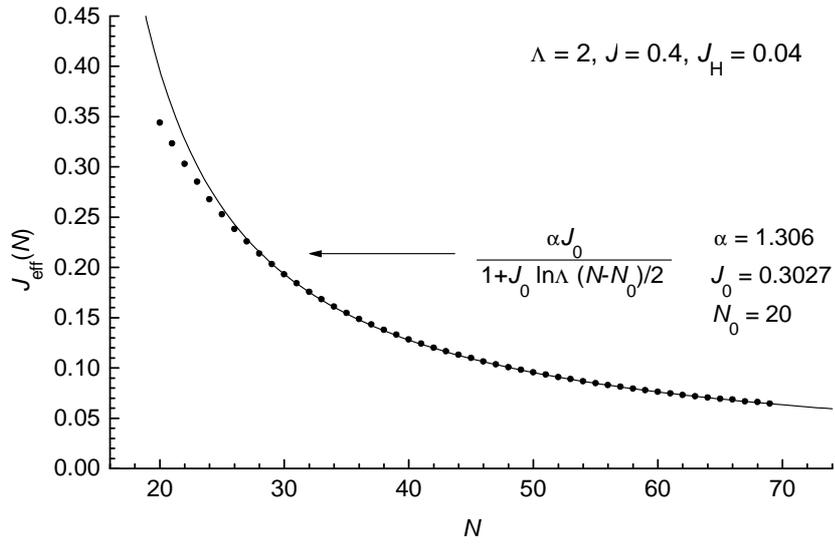
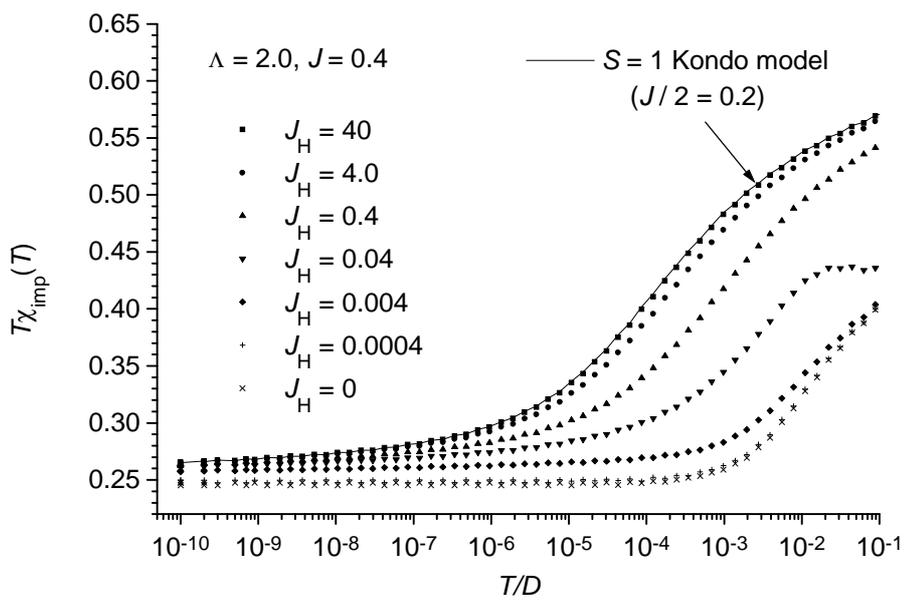
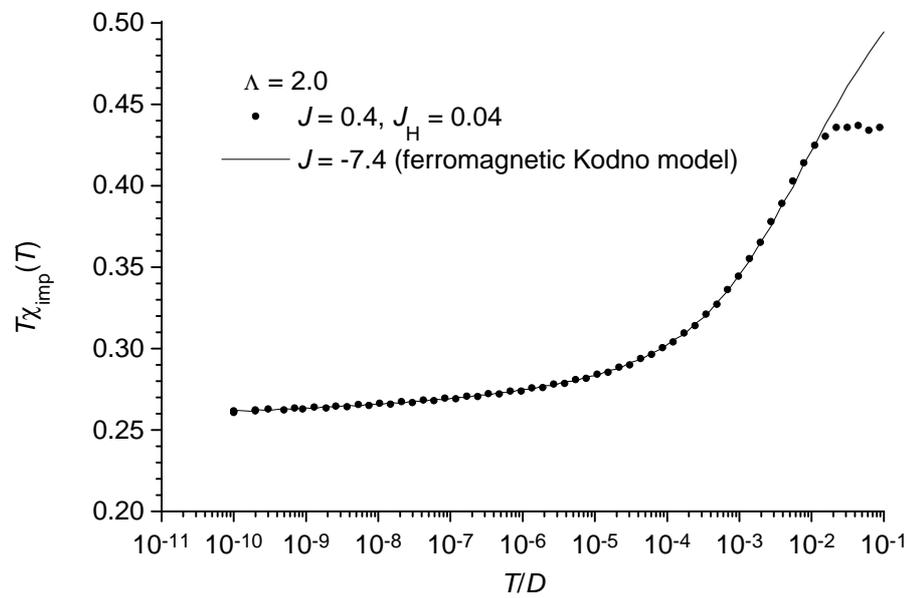
\end{document}